\documentclass[epj]{svjour}
\usepackage{latexsym}
\usepackage{graphics}
\begin{document}
\title{Dynamical Invariants in the Deterministic
Fixed-Energy Sandpile}
\author{Mario Casartelli \inst{1}\inst{2} \and  Luca Dall'Asta\inst{3} \and
Alessandro Vezzani \inst{1} \and Pierpaolo Vivo\inst{4}% etc
}                     % Do not remove
\institute{Dipartimento di Fisica and  CNR - INFM, Universit\`a di
Parma,  Parco Area Scienze 7a, 43100 Parma Italy \and INFN, gruppo collegato di Parma \and Laboratoire de
Physique Th\'eorique,  Batiment 210, Universit\'e de Paris-Sud,
91405 ORSAY Cedex France \and School of Information Systems,
Computing and Mathematics, Brunel University, Uxbridge, Middlesex
UB8 3PH UK}  
\date{Received: date / Revised version: date}
\abstract{
The non-ergodic behavior of the deterministic Fixed Energy Sandpile (DFES), with  Bak-Tang-Wiesenfeld (BTW) rule, is explained by the complete characterization of a class of dynamical invariants (or toppling invariants). The link between such constants of motion and the discrete Laplacian's properties on graphs is algebraically and numerically clarified. In particular, it is possible to build up an explicit algorithm determining the complete set of independent toppling invariants. The partition of the configuration space into dynamically invariant sets, and the further refinement of such a partition into basins of attraction for orbits, are also studied.  The total number of invariant sets equals the graph's complexity. In the case of two dimensional lattices, it is possible to estimate a very regular exponential growth of this number vs. the size. Looking at other features, the toppling invariants exhibit a highly irregular behavior. The usual constraint on the energy positiveness introduces a transition in the frozen phase. In correspondence to this transition, a dynamical crossover related to the halting times is observed.  The analysis of the configuration space shows that the  DFES has a different structure with respect to dissipative BTW and stochastic sandpiles models, supporting the conjecture that it lies in a distinct class of universality. 
}

\PACS{
      {PACS-key}{describing text of that key}   \and
      {PACS-key}{describing text of that key}
     } % end of PACS codes

\authorrunning{Casartelli et al.}
\titlerunning{Dynamical invariants in the Deterministic
Fixed-Energy Sandpile}
\maketitle
\section{Introduction}
Sandpile models have been introduced in statistical mechanics as prototypes for 
the Self-Organized Criticality (SOC), a concept that has been widely used in order 
to explain the appearance of power law correlations in non-equilibrium steady states 
of self-organizing systems with many degrees of freedom~\cite{bak,jensen}.
As conceived by Bak, Tang, and Wiesenfeld (BTW) in their seminal work~\cite{btw}, 
dissipation and external input of grains should be necessary conditions for the 
existence of a self-organized critical state. On the other hand, the common numerical 
technique used to study the critical behavior of these models is the finite-size scaling, 
that requires the analysis of larger and larger systems~\cite{finitesize}.
Increasing the system's size implies that, at the criticality, larger and larger 
avalanches are produced, i.e. the system's self-sustained activity persists in time. 
Since new grains are added only when the dynamics eventually stops (at least in the 
original BTW model~\cite{btw}), and dissipation is localized at the boundaries, 
it has been conjectured that the correct infinite size behavior should be well 
reproduced by models in which both the external drift and the dissipation approach 
zero~\cite{rate}. For these reasons, a new class of conservative sandpile models have 
been introduced, in which dissipation is prevented by periodic boundary conditions and no external input of grains is allowed. These models are called Fixed-Energy Sandpiles (FES)~\cite{fes}, since the total number of grains, or {\em energy}, is a constant of motion, fixed by the initial conditions.
As a consequence, the activity of the system only depends on the total energy. For stochastic (i.e. Manna-like) updating rules, a threshold energy exists, above which the system does not relax to a stable state and a non-zero activity is dynamically  maintained.
Using finite-size scaling techniques, the critical behavior of the stochastic FES has been shown to belong to a particular class of absorbing state phase transitions (APT), different to the directed percolation~\cite{vespignani,sven,dickman,maslov}.
On the contrary, statistical mechanics approaches miss to pinpoint the behavior of deterministic FESs, with BTW updating rule~\cite{btw}.
In fact, BTW deterministic FESs (DFES) present very strong non-ergodic features due to the existence of periodic orbits in which the system eventually enters in all the range of possible values for the total energy~\cite{vespignani,bagnoli}.
In particular, in \cite{bagnoli} DFES has been studied on a square lattice with periodic boundary conditions (discrete torus) and it has been established that by varying the order parameter
(the energy density) the system undergoes a transition from the frozen 
phase with an absorbing state to the active phase  of eventually periodic
dynamics. Moreover, in the active phase, a ``devil staircase'' plateaus structure 
(strictly related to the behavior of the average periods) has been observed.
The recently proposed exact solution for the dynamics of the one-dimensional DFES~\cite{dallasta} corroborates the idea that deterministic BTW model couldn't be comprised into the same universality classes of stochastic or dissipative models.

The present work deals with DFES defined on an unoriented
connected graph. It is focused on the relevance of dynamical invariants 
that are responsible for its non-ergodic features and undermine a purely statistical approach.
We note that a set $\Phi_1,... , \Phi_m$ of distinct constants of motion,
ranging into $v_1,..,v_m$ possible values respectively, determines a partition of
the configuration space ${\cal C}$ into
$ \mathcal{N} = \prod_1^m v_k$ dynamically invariant classes of configurations
(or ``atoms''~\cite{atomi} in the language of dynamical systems). 
Such atoms are, in other terms, the
counterimages of possible values assigned to the invariants.
A natural problem is then the determination 
of the maximal (i.e. the most refined) partition induced by such 
invariants. 

Dynamical invariants will be studied by recovering some important 
results already shown to hold  for the dissipative model \cite{ruelle}. 
In that case, with open boundaries
and random addition of sand grains, invariants cannot play the same 
partitioning role. However, some of our problems  are implicitly 
shadowed there, and most of the algebraic tools may be resumed 
and adapted to our model. In particular Ref. \cite{ruelle} suggests
that dynamical toppling invariants can be generated from the harmonic 
functions of the discrete Laplacian operator.
The Laplacian can be defined on graphs by standard techniques of algebraic 
graph theory \cite{biggs}. By means of modular algebra, we introduce a computational 
technique allowing for the evaluation of the harmonic functions even for 
quite large sizes.

The group theoretical approach proposed by Dhar \cite{ruelle}  has never been attempted 
in the conservative model. Indeed the Abelian Sandpile Group (ASG) is based on two 
dynamical properties: the addition operation and the existence of a unique stable state for the subsequent relaxation (see Refs.~\cite{exactly,redig} for excellent reviews). 
In the conservative model, there is no addition of grains and the final state, being a 
periodic orbit, is not univocally defined. A crucial point of the paper is the  definition  of the Toppling Group of  ``isoinvariant transformations'', 
that can be view as the natural extension of ASG for conservative models. 
Such an algebraic formulation, together with a relevant result of graph 
theory \cite{bacher}, proves that the total number $\mathcal{N}(\mathcal{G})$ of 
invariant atoms generated by modular harmonic functions is related to a remarkable topological property of the graph. Indeed $\mathcal{N}(\mathcal{G})$ 
equals the ``graph complexity''. This quantity, defined as the number of the spanning trees of the graph, has been used in several contexts, such as the study of electrical networks going back to the seminal works by Kirchoff \cite{kirk}.. The graph complexity  can be evaluated by the product of the non-zero eigenvalues of the Laplacian matrix.
More in general, the relation between physical 
properties and topology on graphs is a very interesting issue \cite{cassi}.
In the dissipative models, for instance, the use of conformal field theory has pointed out 
the relevance of boundary conditions in the computation of  the critical exponents of correlation functions~\cite{cft}.

Afterwords, for the relevant case of a $L\times L$ torus, \cite{bagnoli} we shall compute
the complete class of independent invariants for size $L<24$ pointing out that  the distribution of the invariants  vs. $L$ has a highly complex and unpredictable structure.
In the meantime, the total number of invariant atoms has a very regular behavior 
which can be also analytically evaluated within the general
framework of graph theory. In particular, the number of atoms grows exponentially
with the size of the system. The large number of toppling invariants provides a qualitative description of the non ergodic behavior of the DFES. However such an explanation is not complete, since the basins of attraction (sets of configurations evolving into a periodic orbit or an absorbing state) are contained in the atoms, giving rise to a subpartition. By means of  ``isoinvariant transformations'' we estimate the number of basins per atoms. Interestingly, such estimates provide 
a link with the ``plateaus scenario'' presented in \cite{bagnoli}.
The centers of the plateaus correspond indeed to peaks of the number of basins per atom.
At least for small sizes, the calculated explicit form of the invariants allows for a direct exploration of the internal structure of the partition.
Imposing the usual physical constraint of energy positivity
at each site, we get that at low energies most of the atoms are empty; while at high 
energies all of them are filled by more or less the same number of 
configurations. As a consequence, a transition point $\bar{E}_1$ between these
different regimes can be evaluated. 
We show that there exists a dynamical counterpart of such a transition, consisting
in a qualitative change at $\bar{ E}_1$ of the only relevant dynamical observables
for a frozen system, i.e. the halting time and its fluctuations.

The plan of the paper is the following. In Section~\ref{sec2}, we
define the DFES model with BTW
dynamics for a general unoriented graph $\mathcal{G}$. In
Section~\ref{sec3} the class of independent dynamical
invariants generated by the discrete Laplacian is defined.
Section~\ref{sec4} is devoted to the extension to DFES of the algebraic approach 
of \cite{ruelle} by means of ``isoinvariant transformation'' and Toppling Group.
In Section~\ref{sec5}, we compute the independent invariants for tori of size $L<24$. 
A numerical study of the
refinement of invariant atoms into basins is presented in Section~\ref{sec6}. 
In Section~\ref{sec7}, a detailed study of the structure of 
atoms partitions is provided for small size ($L=4$). Finally in 
Section~\ref{sec8} the low energy transition induced by energy
constraints is discussed. 
Conclusions and outlook for the future work are presented in Section~\ref{sec9}. 
The proofs of the theorems, a brief survey of modular algebra and the
numerical algorithm used to find independent invariants are provided
in the Appendices.

\section{The model: definition and properties}\label{sec2}

Let us consider a generic connected unoriented graph $\mathcal{G}$ with $N$ vertices, 
labeled by integer $i=1, 2, \dots, N$, connected pairwise by a set of unoriented edges
defining a neighbouring relation $i \sim j$.
At each site $i$, we introduce an integer variable $z(i) \in \mathbf{Z}$, 
the number of sand grains in the site, physically interpreted as a local amount of 
{\em energy}. 
A configuration $Z$ is the observable $Z: \mathcal{G} \to {\cal C}$, where the configuration space ${\cal C} = {\bf Z}^{N}$ is the $N$-dimensional domain of integers.
Moreover, each site is endowed with a fixed integer parameter, the {\em critical energy} $z^{c}(i)$, such that if
$z(i) > z^{c}(i)$, the site $i$ becomes {\em metacritical} and $d_i$ units of energy are equally redistributed among its neighboring sites, where the degree $d_i$ of the vertex $i$ is the number of its neighbours. Such redistribution event is called {\em toppling}.
The BTW dynamics consists of a parallel updating, in discrete time, of all metacritical states; therefore the evolution rule for $Z_{t} = \{ z_{t}(i)\}$ can be summarized as follows,
\begin{equation}
\label{evo}
z_{t+1}(i) =z_{t}(i)- \sum_{j} \nabla^{2}_{ij} \theta{\left[z_{t}(j)-z^{c}(j)\right]}~,
\end{equation}
where $\theta(x) = 1 (0)$ if $x > 0$ ($x \leq 0$) and the integer $N\times N$ matrix $\nabla^{2}$ is the Laplacian matrix \cite{biggs} 
\begin{eqnarray}
\nabla^{2}_{ij} = \left\{
\begin{array}{l l}
  d_{i} & \quad \mbox{if $j=i$}\\
  -1 & \quad \mbox{if $i \sim j$}\\
  0 & \quad \mbox{otherwise}~,\\
  \end{array} \right.
\label{delta}
\end{eqnarray}
In the following, the  evolution rule 
(\ref{evo}) will be indicated with the operator $ \mathcal{U}_{BTW} $, 
i.e. $Z_{t+1}=\mathcal{U}_{BTW} Z_{t}$.
If the total energy $E=\sum_i z(i)$ is sufficiently high, topplings propagate, creating {\it avalanches} of energy that  cover the whole system.
It is a common convention to assume the site variables bounded between $0$ and a maximal value $i_{max}=2d_{i}-1 $. The lower bound is simply due to the fact that negative amounts of grains are unphysical, while the upper one tends to avoid the useless treatment of energy ranges forcing all sites to be active. Here, we consider the whole space ${\bf Z}^{N}$ for a general approach, pointing out where necessary the effects of such additional constraints. The forward motion is completely deterministic, and the system eventually falls into a closed orbit, i.e. a fixed point or a limit cycle. Orbits $\{Z_0, Z_1, Z_2,... \} $ in ${\cal C}$ are thus completely determined by the initial condition $Z_0$.

An important remark concerns the symmetry invariances of the dynamics. First the BTW toppling redistribution rule is completely symmetric, thus, given two configurations $Z$ and $Z'$, if there exists a transformation $\xi$ (belonging to the symmetry group of the graph) such that $\xi Z = Z'$ and $\xi^{-1} Z' = Z$, then $\xi$ commutes with the dynamics ($\xi \circ \mathcal{U}_{BTW}$ $=$ $\mathcal{U}_{BTW}\circ \xi$). The second invariance is due to an internal symmetry of the evolution rule with respect to the transformation $z(i)$ $\to$ $z'(i)$ = $2d_{i}-1 - z(i)$ $\forall i$.

Recently, Bagnoli and coworkers~\cite{bagnoli} have investigated numerically the activity patterns of the DFES model on  two-dimensional tori of different linear sizes $L$.
Varying the energy density $\bar{E}=E/L^2$, the system undergoes a transition from a frozen phase (absorbing state) to the active phase characterized by eventually periodic dynamics. Moreover the phase diagram (the density of active , i.e. metacritical, sites vs. $\bar{E}$),
displays a plateaus structure organized as a devil-staircase~\cite{bagnoli}.
In correspondence to these plateaus, the average length of the periods is small 
and approximately constant for increasing sizes $L$.
A similar step-like structure with very short periods is partially recovered in 
the one-dimensional model, for which an exact solution is presented in Ref.\cite{dallasta}. 
Some of the intriguing dynamical features of the two dimensional case will be 
qualitatively explained in Section~\ref{sec6}.

\section{Toppling invariants and discrete harmonicity}\label{sec3}

The general concept of ``toppling invariant'' for a sandpile
model on any kind of topology is quite obvious.
Let  $Z_t$ and  $Z_{t+1}$ denote consecutive configurations along the orbit started from
$Z_0$. A functional $\Phi(Z)$  may be defined a \textit{toppling
invariant} if, for every $t$,
\begin{equation}\label{inva}
\Phi (Z_{t+1})= \Phi (Z_t)~.
\end{equation}

The simplest conserved scalar form on ${\bf Z}^{N}$ is the
total energy, that suggests to use linear integer functions of the site energies.
On a periodic one-dimensional lattice the exact solution 
shows that  constant of motions can be defined by means of 
linear scalar functions $\Phi$ that are invariant modulo the lattice size $N$.
Such a property is a consequence of the simple linear periodic geometry; then,
on a general graph, we should rather consider functions through a ($\bmod K$)-linear form.

On a generic connected unoriented graph $\mathcal{G}$ (without multi-links and self-links),
we introduce a set ${\cal{F}}(\mathcal{G},K)$ of functions
$f:~ \mathcal{G} \to{\bf Z}_K$ defined on the nodes of the graph $\mathcal{G}$ and taking values in the ring ${\bf Z}_K =\{0,1,\dots,K-2,K-1\}$.
A $K\mathcal{G}$\textit{-functional} $\Phi_f(K,\mathcal{G},Z)$
generated by ${f} \in {\cal{F}}(\mathcal{G},K)$  is the quantity
 \begin{equation}
 \label{functional}
 \Phi_f(K,\mathcal{G},Z) = \left( \sum_{i=1}^{N}
{f}(i)~z(i)~\right) {\bmod} K~.
\end{equation}

The range of a $K\mathcal{G}$-functional is between $0$ and $K-1$.
Furthermore, the restriction $f(i)\in{\bf Z}_K$
is not relevant. Indeed, a substitution in Equation
(\ref{functional})of  $f\in{\cal{F}}(\mathcal{G},K)$  with 
$f':~ \mathcal{G} \to{\bf Z}$
and $f(i)= f'(i)~ {\rm mod}~K $, gives rise exactly to the same functional.
\def\=K{\mathop{=}\limits^K}
\def\pka{\mathop{=}\limits^{p^{\kappa}}}
\def\pio{\mathop{=}\limits^{p^{\iota}}}
\def\pie{\mathop{=}\limits^{p^{\epsilon}}}
\def\ph{\mathop{=}\limits^{p^{h}}}

Hereafter, we shall use the notation $\=K$ for the equalities between elements of
${{\bf Z}}_K$. In such expressions, any integer has to be
replaced with its $K$-modulo and all operations
are to be intended in this ``modular'' sense.
The space ${\cal{F}}(\mathcal{G},K)$, endowed with the  ${\rm mod}~ K$ sum, is
a finite Abelian group, and therefore we will use the usual  definitions
of finite group theory, such as element periodicity \cite{nota}, generator set,
and group morphism. Relevant information on the $K$-(Abelian)modulo 
structure of ${\cal{F}}(\mathcal{G},K)$ \cite{beachy} are summarized
in Appendix \ref{appen1}.

Let us introduce the group morphism  ${\widetilde{\nabla}_K}^2:{\cal{F}}(\mathcal{G},K)\to{\cal{F}}(\mathcal{G},K)$
 which is the natural realization in ${\cal{F}}(\mathcal{G},K)$ of the 
usual Laplacian operator defined by the matrix $\nabla^2$. In particular,
for functions $f\in {\cal{F}}(\mathcal{G},K)$,
${\widetilde{\nabla}_K}^2$, is defined by
 \begin{eqnarray}
  (\widetilde{{\nabla}}^2_K f)(i) &:\=K (\nabla ^2 f)(i)~.
\label{dlap}
\end{eqnarray}

  A  function $h\in {\cal{F}}(\mathcal{G},K)$
is said to be (discrete) $K$-\textit{harmonic} on $\mathcal{G}$ if
 \begin{equation}
 \label{dKh}
 ({{\nabla}}^2 h)(i) \=K 0 ~~~~ \forall i \in \mathcal{G}~.
\end{equation}
Alternatively, $h$ is $K$-harmonic on $\mathcal{G}$ if it belongs to the kernel of
${\widetilde{\nabla}_K}^2$. Such ${\rm Ker}({\widetilde{\nabla}_K}^2)$ is a subgroup
of ${\cal{F}}(\mathcal{G},K)$ and it will be denoted by ${\cal{H}}(\mathcal{G},K)$.

Functionals generated by $K$-harmonic functions play a relevant role in  FES
dynamics. In particular, the $K\mathcal{G}$-functional $\Phi_f(K,\mathcal{G},Z_t)$ is a toppling invariant if and only if its generating function $f$ is $K$-harmonic in ${\cal G}$.
The proof of this result is provided in Theorem 1 of Appendix \ref{appen2}.

The energy, generated by the constant harmonic function $c=1$, is the only
$K$-harmonic function for any value of $K$.
Fixing the energy is equivalent to define the invariant ``energy surface''
where the motion takes place.
In analogy with the conserved quantities (motion invariants) of classical
mechanics, Theorem 1 implies that the configuration space is divided
into dynamically separated subsets, characterized by the value of $\Phi_f(K,{\cal G},Z)$.
In other terms, these invariants establish a {\em partition} of the energy surface,
i.e. an exhaustive collection of disjoint subsets, that in the dynamical systems theory are
called {\it atoms} \cite{atomi}.
It is worthy noting that not all the $K$-harmonic functions are independent.
For example, for $f,g\in {\cal{H}}({\cal G},K)$, if  $f$ and $g\=K 2f$ 
then $\Phi_g(K,{\cal G},Z)\=K 2 \Phi_f(K,{\cal G},Z)$.
Therefore, for each configuration $Z$, the value of  $\Phi_{g}(K,{\cal G},Z)$ is
determined by $\Phi_f(K,{\cal G},Z)$.

In general, the function  $h\in{\cal{H}}({\cal G},K)$ is defined to be dependent 
on the functions $h_n\in{\cal{H}}({\cal G},K_n)$ if there exists a set of numbers $a_j$
such that for any function $Z$, one has
\begin{equation}
 \label{dep_inv}
  \Phi_h(K,{\cal G},Z) \=K a_0 E(Z) + \sum_{j=1}^N a_j \Phi_{h_j}(K_j,{\cal G},Z)
\end{equation}
The energy of the system $E(Z)$ has been considered apart,
so that the independent invariants can be defined neglecting an arbitrary constant.

For instance, in a one-dimensional system the enumeration of all the independent functions 
is trivial; the structure of the discrete Laplacian imposes $N-2$ constraints on $N$ variables. 
The two independent functions are for instance those corresponding to the total energy  and the 
``linear momentum'' ${\rm mod} N$ along the periodic system ($h_{2}(i)=i, \forall i$).
Increasing the dimensionality of the lattice or changing the topology to a generic graph, 
the structure of the configuration space becomes more and more complicated and the 
number of dynamical invariants rapidly grows with the system size.

In general, a complete set of independent dynamical invariants can be computed by
the Smith decomposition of the Laplacian matrix $\nabla^2$ (see Ref.~\cite{ruelle}) as 
\begin{equation}
\label{smith}
S=A \nabla^{2} B = {\it diag}(0,g_{1},g_{2},\dots,g_{N-1})~,
\end{equation}
where $A$ and $B$ are integer matrices with determinant $\pm 1$ (unimodolar matrices)
The $k-$th column of $B$ in (\ref{smith}) defines an independent harmonic function
 ${\rm mod}~g_k$. We have two important corollaries of this result:

\noindent 1 -  the total number
of independent invariants is finite;

\noindent 2 - the number of atoms defined
by such non constant invariants is 
${\cal N}({\cal G})=\prod_{k=1}^{N-1} g_k$.

An explicit calculation via Smith decomposition is possible only for graphs of small
sizes, because, for intrinsic features of the algorithm, integer numbers
greater than the maximum allowed by the computer are early involved.
For example, on  $L\times L$ tori, the transformation (\ref{smith}) may be worked out
only up to $L=5$. An alternative approach, described in details
in Appendices C and D, allows to compute the invariants for tori 
with $L$ up to $24$. Unfortunately, this alternative method cannot assure
the completeness of the set of invariants. In our experiments, completeness is
guaranteed only for $K < K_{\mathcal{G}} = 20000$. This is 
once again a computational bound. More in general, for a graph $\cal G$ and 
a suitable integer  $ K_{\mathcal{G}} $ the algorithm
provides a set $I_{{\cal G},K_{\cal G}}$ of independent functions
$h_j\in{\cal{H}}({\cal G},p_j^{\iota_j})$ of periodicity $p_j^{\iota_j}$
($p_j$ are primes, $\iota_j$ are integers and
$p_{j}^{\iota_j} < K_{\cal G}$ $\forall j$). Furthermore, every function 
$h \in{\cal{H}}({\cal G},K)$  ($K =\prod_{j=1}^{j_{max}} {\theta_j}$ 
with ${\theta_j} \in {\bf N}$, $\theta_j <K_{\cal G}$) 
depends on the set $I_{{\cal G},K_{\cal G}}$.

\section{Toppling Group and Isoinvariant Transformations}\label{sec4}

On the configuration space it is possible to introduce a class of
transformations preserving the values assumed by the constants of motion.
In particular, an ``isoinvariant transformation'' is a mapping $\eta$: ~ $Z \in {\cal C}$
$\to Z' \in {\cal C}$
such that $\Phi_h(K,{\cal G},Z)=\Phi_h(K,{\cal G},Z')$ for any integer $K$
and any $K$-harmonic function $h$.

Theorem 2, reported in Appendix \ref{appen2}, states that a 
necessary and sufficient condition for a transformation to be
isoinvariant is that it can be written in the form
\begin{equation}
Z' = \eta(Z) = Z + {\nabla}^2 U
\label{iso}
\end{equation}
where $U \in {\bf Z}^{N}$ is an integer vector.

We observe that the class ${\{ \eta \}}_{iso}$ of transformations, defined
by (\ref{iso}), operating linearly on the configuration space, constitutes
an Abelian group.
A set of generators for this group are the elementary toppling operators
$\eta_{i}$, decreasing by $d_{i}$ grains the site $i$, and augmenting all its neighbors by $1$ unit. Hence such group will be called 
the Toppling Group.
The BTW dynamics itself is an example of isoinvariant transformation,
depending however at each step on the metacritical domain of the state.
This makes in general $\eta {\mathcal U}_{BTW}Z\not= {\mathcal U}_{BTW} \eta Z$.
(Of course, in a reduced configuration space ${\cal C'}$, we should admit only those
transformations ensuring that $Z'$ belongs to it).

Some global properties of the toppling group  can be studied by
means of  algebraic graph theory (see for instance Refs. \cite{biggs}).
Particularly important is the concept of {\em preflow} that has been
introduced in \cite{bacher}.

Given a graph $\mathcal{G}$, we call integer ``preflow'' an integer-valued
function defined on its edges after having assigned them an orientation.
The preflows constitutes a group with the addition in ${\bf Z}$.

The isoinvariant transformations are a subgroup of the preflows on
$\mathcal{G}$. In our context, we can apply a relevant
result on integer preflows stated in Ref. \cite{bacher}. More precisely, in
Appendix \ref{appen5} we will prove that the number ${\cal N}({\cal G})$ of atoms
generated by toppling invariants of the form (\ref{functional}) equals the
{\it complexity} of the graph $\mathcal{G}$, i.e. the number of its
spanning trees \cite{biggs}. For example, in a generic tree-graph the  number of spanning trees is 1, while on a chain is the number of sites.  
Moreover, this number can be evaluated from the Laplacian matrix $\nabla^{2}$ as the
product of all non-zero eigenvalues $\lambda_{i}$ , i.e.
\begin{equation}\label{vol}
{\cal N}({\cal G}) = \frac{1}{N} \prod_{i=2}^{N} {\lambda}_{i}~.
\end{equation}
\noindent (It is a classical result of algebraic graph theory that the 
quantity (\ref{vol}) is an integer).

A sort of partition into atoms and isoinvariant transformations,
however with different meaning and physical interpretations, 
were already defined in Dhar's algebraic approach \cite{ruelle,exactly}.
In the dissipative model, partitions do not correspond to constants
of motion (the system remains within an atom 
during an avalanche, jumping to different atoms by addition of new grains).
Another relevant property of dissipative  
model is that in each atom there is one and only one recurrent configuration,
this property does not have a counterpart in FES, where recurrent configurations
cannot be defined. 
 Moreover, from an algebraic point of view, we note that in the
open case, in order to take into account dissipation, the matrix playing the
role of the Laplacian $\nabla^2$ is not singular, permitting a set of operations
not allowed in our case (e.g. the definition of atoms 
using the inverse of the matrix). Anyway,  in \cite{exactly}  the 
link with the graph complexity is also put into evidence following a
different path, the so called ``burning test'', instead of our preflows 
criterion.

\section{Invariants on a torus}\label{sec5}

In this Section we focus on the case in which ${\cal G}$ is a 
two-dimensional regular periodic lattice of size $N=L^2$, i.e. a torus; 
in this system, the non-ergodic properties of FES dynamics was
observed for the first time in ~\cite{fes,bagnoli}.
The toppling invariants partition will be studied by means of the algorithm
introduced in the appendices \ref{appen3} and \ref{appen4}. 
By considering $K_{\cal G}=20000$ 
(the bound $K_{\cal G}$ is introduced in Section~\ref{sec3}),
we are able to compute the 
invariants for sizes $L$ up to 24.
The results for $L\leq 13$ are reported in Table \ref{tab1}.
For such sizes the algorithm seems to be  efficient to find a complete 
set of invariants; indeed, the largest periodicity appearing in Table 
\ref{tab1} is much smaller than $K_{\cal G}=20000$. For $L\leq 5$ this has 
been confirmed by a direct evaluation of the whole set of independent 
invariants by means of a Smith decomposition (as already pointed out, 
the Smith decomposition cannot be easily applied to larger systems). 
In Table \ref{tab1} we call $Q_{p^\iota}$ the number of functions of periodicity ${p^\iota}$
in the class $I_{{\cal G},K_{\cal G}}$ of invariants.
\begin{table}[t]
\begin{center}
$\begin{array}{|c|c|c|c|c|c|c|c|c|c|c|c|}
\hline
\hline
  L  & \multicolumn{11}{|c|}{{\rm Module }~p^{\iota}} \\ \cline{2-12}
     & \multicolumn{11}{|c|}{ Q_{p^\iota}} \\
\hline \hline
  3  &  2  & 3  & 3^2 & ~ & ~ & ~ & ~& ~ & ~ & ~ & ~\\ \cline{2-12}
  ~  &  4  & 2  &  2  & ~ & ~ & ~ & ~& ~ & ~ & ~ & ~\\ \hline
  4  &  2  & 2^3  & 2^5 & 3  & ~ & ~ & ~& ~ & ~ & ~ & ~\\ \cline{2-12}
  ~  &  2  & 4  &  1 & 4 & ~ & ~ & ~& ~ & ~ & ~ & ~\\ \hline
  5  &  2  & 5  & 5^2 & ~ & ~ & ~ & ~& ~ & ~ & ~ & ~\\ \cline{2-12}
  ~  &  8  & 2  &  6  & ~ & ~ & ~ & ~& ~ & ~ & ~ & ~\\ \hline
  6  &  2  & 2^3  & 3 & 3^2  & 5 & 7 & ~& ~ & ~ & ~ & ~\\ \cline{2-12}
  ~  &  2  & 9  &  2 & 2 & 4 & 4 & ~& ~ & ~ & ~ & ~\\ \hline
  7  &  2  & 7  & 7^2 & 13  & ~ & ~ & ~& ~ & ~ & ~ & ~\\ \cline{2-12}
  ~  &  12  & 10  &  2 & 8 & ~ & ~& ~ & ~ & ~ & ~ & ~\\ \hline
  8  &  2  & 2^3  & 2^5 & 2^7  & 3 & 7 & 17& ~ & ~ & ~ & ~\\ \cline{2-12}
  ~  &  2  & 4  &  8 & 1 & 4 & 8 & 4& ~ & ~ & ~ & ~\\ \hline
  9  &  2  & 3  & 3^2 & 3^3  & 3^4 & 17 & 37& ~ & ~ & ~ & ~\\ \cline{2-12}
  ~  &  16  & 10  &  2 & 2 & 2 & 8 & 8& ~ & ~ & ~ & ~\\ \hline
  10  &  2  & 2^3  & 3 & 5  & 5^2 & 11 & 29 & 41 & ~ & ~ & ~\\ \cline{2-12}
  ~  &  2  & 17  &  8 & 2 & 6 & 8 & 4& 4 & ~ & ~ & ~\\ \hline
  11  &  2  & 11  & 11^2 & 89  & 109 & ~ & ~ & ~ & ~ & ~ & ~\\ \cline{2-12}
  ~  &  20  & 18  &  2 & 8 & 8 & ~ & ~& ~ & ~ & ~ & ~\\ \hline
  12  &  2  & 2^3  & 2^4 & 2^5  & 3 & 3^2 & 3^3 & 5 & 7 & 11 & 13\\ \cline{2-12}
  ~  &  2  & 4  &  16 & 1 & 6 & 6 & 6 & 12 & 4 & 12 & 8\\ \hline
  13  &  2  & 5  & 13 & 13^2  & 233 & 313 & ~& ~ & ~ & ~ & ~\\ \cline{2-12}
  ~  &  24  & 8  &  18 & 6 & 8 & 8 & ~& ~ & ~ & ~ & ~\\ \hline
\end{array}$
\end{center}
\caption{\small Number $Q_{p^{\iota}}$ of independent harmonic functions of periodicity $p^{\iota}$ for $L$ between $2$ and $13$ and $K_L = 20000$.}
\label{tab1}
 \end{table}

\begin{figure}
\begin{center}
\resizebox{0.75\columnwidth}{!}{
\includegraphics{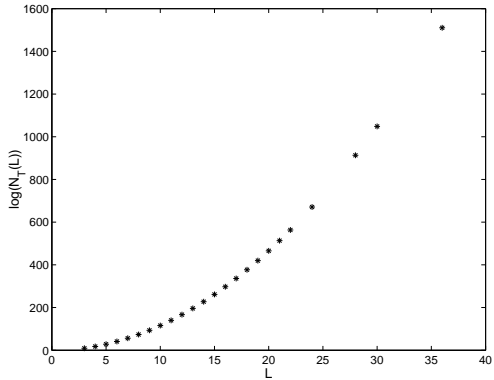}
}
\label{inv}
\caption{The natural logarithm of the total number of  atoms $\log({\cal N}({\cal G}))$
as a function of the torus size $L$.}
\end{center}
\end{figure}

Table (\ref{tab1}) evidences that the number of invariant 
as functions of $L$, $p^{\iota}$ and $Q_{p^\iota}$
is very irregular and it seems that no simple rule
can be inferred to foresight results corresponding to larger sizes.
On the contrary, the total number of atoms ${\cal N}({\cal G})$ seems 
to have a quite regular behavior. Indeed, ${\cal N}({\cal G})$
can be evaluated as
\begin{equation}
{\cal N}({\cal G})=\prod_{p^\iota<K_{\cal G}} p^{\iota \cdot  Q_{p^\iota}}~.
\label{numberpk}
\end{equation}
We exploit the property that an invariant generated by a $K$-periodic function  
can have $K$ values, and we assume that $I_{{\cal G},K_{\cal G}}$ is a complete set. 
The logarithmic plot of ${\cal N}({\cal G})$ vs. $L$
reveals a clearly quadratic behavior (see Figure \ref{inv}). More precisely
\begin{equation}
{\cal N}({\cal G})\sim \exp(c L^2)
\label{Nas}
\end{equation}
where $c=1.20\pm 0.05$. In the figure all $L\leq24$ and some of the
larger sizes have been considered. For some values of $L$ we get a
different behavior, however it is likely that such anomalies could
depend on invariants of periodicity larger than
$K_{\cal G}=20000$, which are not captured by the actual
implementation of the algorithm.
${\cal N}({\cal G})$ can also be evaluated by means of formula (\ref{vol}) in this case
the non-zero eigenvalues of the Laplacian matrix are analytically known, this gives
\begin{equation}\label{count_inv}
{\cal N}({\cal G}) = \frac{1}{L^2} \prod_{\ell=1}^{L-1} \prod_{m=1}^{L-1} \left[ 4 - 2\cos{\left( \frac{2\pi \ell}{L}\right)} -2\cos{\left( \frac{2\pi m}{L}\right)}\right]~.
\end{equation}
For large $L$, ${\cal N}({\cal G})\sim \exp(c_{th} L^2)$, where the parameter $c_{th}$ can be computed as~\cite{exactly}
\begin{equation}\label{cth}
c_{th}  \simeq \int_{0}^{2\pi}\int_{0}^{2\pi}\frac{d\alpha d\beta}{4\pi^{2}} \log{\left( 4 - 2\cos \alpha - 2\cos \beta \right)} \simeq 1.17,
\end{equation}
that confirms the results obtained by means of numerical evaluations.

\section{Atoms and basins}\label{sec6}

The exponential growth of the atoms for the DFES on the torus provides a qualitative
explanation of its non ergodic features, since it implies a
fragmentation of the configuration space into a very large number of dynamically
intransitive domains. However, such an invariants-based partition does not
resolve completely the non ergodicity. Dynamical basins of attraction determine indeed 
a further non trivial subpartition, whose properties will be investigated in this section by means of isoinvariant transformations introduced in Sec.4. By iterating such
transformations, one can generate indeed a sufficient number of states for a
reliable estimate of the basin weights within every single atom, revealing its
internal structure.
\begin{table}[t]
\begin{center}
$\begin{array}{|c|c|c|c|c|c|}
\hline
\hline
  {\rm  Number~of}   & \multicolumn{5}{|c|}{{\rm Orbit~period~} \ell} \\ \cline{2-6}
  {\rm basins}  & \multicolumn{5}{|c|}{{\rm Basins~of~period~} \ell} \\
\hline \hline
  ~  &  118  & ~  & ~  & ~ & ~ \\ \cline{2-6}
  8  &   8   & ~  & ~  & ~ & ~ \\ \hline \hline
  ~  &  8  & 16  & ~  & ~ & ~ \\ \cline{2-6}
  749  &  405  & 344  & ~  & ~ & ~ \\ \hline \hline
  ~  &  85  & 170  & ~  & ~ & ~ \\ \cline{2-6}
  81  &   2   & 79  & ~  & ~ & ~ \\ \hline \hline
  ~  & 8   & 16  & ~  & ~ & ~ \\ \cline{2-6}
  817  & 782   &  35  & ~  & ~ & ~ \\ \hline \hline
  ~  &  74  & ~  & ~  & ~ & ~ \\ \cline{2-6}
  1  &   1   & ~  & ~  & ~ & ~ \\ \hline \hline
  ~  &  8  & 16  & 24  & ~ & ~ \\ \cline{2-6}
  852  &   364   & 1  & 487  & ~ & ~ \\ \hline \hline
  ~  &  50  & ~  & ~  & ~ & ~ \\ \cline{2-6}
  4  &   4   & ~  & ~  & ~ & ~ \\ \hline \hline
  ~  &  8  & 16  & 32  & 24 & 96 \\ \cline{2-6}
  822  &  183   & 46  & 524  & 16 & 53 \\ \hline \hline
  ~  &  116  & ~  & ~  & ~ & ~ \\ \cline{2-6}
  418  &   418   & ~  & ~  & ~ & ~ \\ \hline \hline
\end{array}$
\end{center}
\caption{Some typical atoms partitioning in different basins of attractions at the energy density $\bar{E}=2.25$.
The first column reports the number of basins contained in the atom. The next columns display how such number is distributed for basins corresponding to orbits of different period length.
The data are obtained numerically averaging over $200$ atoms and $1000$ initial conditions for each atom.}
\label{tab3}
 \end{table}

In particular, for a torus of size $L=40$, at different energy densities,
we collected data for $200$ atoms and, in each atom, $1000$ initial states (obtained by 
extracting random integer vectors for the transformation (\ref{iso})).
Some typical results for an energy density $\bar{E}=2.25$ are displayed in Table (\ref{tab3}). 
Each couple of rows corresponds to a different atom. The first column reports the number
of different basins identified over $1000$ initial configurations.
In the next columns basins are grouped according to the period of the final orbit (in the
first row there is the orbit length, in the second the number of configurations evolving
into an orbit of such length). The energy $\bar{E}=2.25$ corresponds to a plateau where the average orbit length is eight (see \cite{bagnoli}). Nevertheless, due to finite size effects,  a wide range of orbit lengths is present. 
Different behaviors can be observed. In particular: 1) there are atoms divided into
very few  (or even one)  basins; 2) there exist atoms divided into many (up to
hundreds) basins. A reasonable property (frequently observed  indeed) is that the former case corresponds to very long orbits, and the latter to short ones. However, there are remarkable exceptions. It is also noteworthy that, when an atom is divided into many basins, only few periods are present, and all of them are simple multiple of a minimal period. This is true for all energies.

\begin{figure}
\begin{center}
\resizebox{0.75\columnwidth}{!}{
\includegraphics{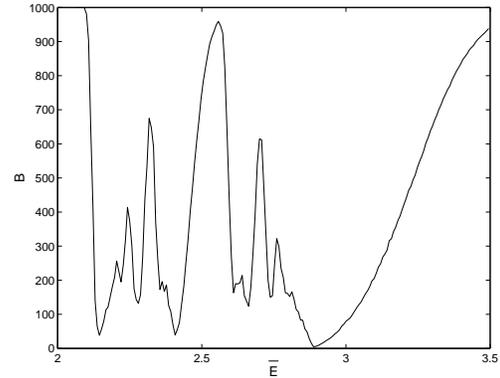}
}
\caption{The average number of different basins, emerging from 1000
states of the same atom, vs. $\bar{E}$. In the average, 200 atoms are considered 
for each energy. This behavior does not depend on the
number of starting configurations or the lattice size.}
\label{basins}
\end{center}
\end{figure}

In Figure \ref{basins}, we plot, as a function of $\bar{E}$, the average number
of distinct basins reached starting from $1000$ configurations belonging to the same atom
(the average is obtained considering $200$ different atoms at the same energy).
We checked that this behavior does not depend on the
number of starting configurations or the lattice size.
Indeed, by varying such parameters, we obtain an analogous figure
after a suitable rescaling of $E$ and $B$.
A comparison with data in  Ref.\cite{bagnoli}
shows that the plateaus characterized by short periods correspond to peaks 
in Figure \ref{basins}. Hence, in the middle of the plateaus the atoms are 
highly fragmented into many basins, while in the transition regions between 
different plateaus there are few basins per atom. This suggests that the  
dynamics of the system plays a non trivial role in the refinement of invariant atoms,
confirming that there is a structural difference with respect 
to the dissipative case: indeed, in the Dhar's algebraic approach, the partition 
into atoms provides a complete description of the configuration space, since
to each atom corresponds a single absorbing (recurrent) state (see Sectio~\ref{sec4}).
The possible existence of further simmetries induced by the deterministic BTW rule
suggests that the DFES model belongs to a different class of universality. Similar
simmetry based arguments may be used to distinguish the DFES model from stochastic
models.

\section{ Atoms filling for two-dimensional lattices of small size}\label{sec7}

Up to now we assumed  that the configuration space ${\cal C}$ is extended 
to the whole ${\bf Z}^{N}$. However, for reasons recalled in Section~\ref{sec2}, 
in sandpile models the energy is generally assumed to be non negative, and it 
is also bounded by a maximum value. 
Hereafter, we assume that on a toroidal lattice $0\leq z_i \leq 7 ~\forall i$, calling this reduced configuration space ${\cal C'} \subseteq {\cal C}$.

A natural problem arising from such constraints is the way atoms are filled by
configurations. Indeed, while in ${\cal C }$ there are infinite
configurations compatible with the invariants, in ${\cal C'}$ their number is finite,
and it depends on ${\bar E}$. Since this number exponentially grows with the size $N=L
\times L$, an exhaustive numerical analysis is bounded to $L=4$ (see Table \ref{tab2} 
for a complete set of invariants in the matricial representation).

\begin{table}[t]
\begin{center}\begin{tabular}{|c|}\hline
~\\
$\left(
\begin{array}{cccc}
0 & 1 & 0 & 0 \\
1 & 0 & 1 & 0\\
0 & 1 & 0 & 0\\
0 & 0 & 0 & 0\\
\end{array}
\right)
\left(
\begin{array}{cccc}
1 & 0 & 0 & 0 \\
0 & 1 & 0 & 1 \\
1 & 0 & 0 & 0 \\
0 & 0 & 0 & 0 \\
\end{array}
\right)
{\rm for}~ p^{\kappa}=2$ \\
~\\
$\left(
\begin{array}{cccc}
4 & 4 & 0 & 7 \\
4 & 4 & 5 & 0\\
0 & 3 & 0 & 0\\
1 & 0 & 0 & 0\\
\end{array}
\right)
\left(
\begin{array}{cccc}
1 & 0 & 3 & 0 \\
4 & 4 & 4 & 4 \\
3 & 0 & 1 & 0 \\
0 & 0 & 0 & 0 \\
\end{array}
\right)
\left(
\begin{array}{cccc}
0 & 4 & 4 & 7\\
5 & 4 & 4 & 0\\
0 & 3 & 0 & 0\\
0 & 0 & 1 & 0\\
\end{array}
\right)
\left(
\begin{array}{cccc}
4 & 0 & 7 & 4 \\
4 & 5 & 0 & 4 \\
3 & 0 & 0 & 0 \\
0 & 0 & 0 & 1 \\
\end{array}
\right)
{\rm for}~ p^{\kappa}=8$ \\
~\\
$\left(
\begin{array}{cccc}
25 & 24 & 29 & 4 \\
8 & 9 & 24 & 25 \\
5 & 12 & 1 & 0 \\
0 & 1 & 0 & 1 \\
\end{array}
\right)
{\rm for}~ p^{\kappa}=32$ \\
~\\
$\left(
\begin{array}{cccc}
1 & 2 & 1 & 0 \\
2 & 2 & 2 & 1 \\
1 & 2 & 1 & 0 \\
0 & 1 & 0 & 0 \\
\end{array}
\right)
\left(
\begin{array}{cccc}
1 & 2 & 1 & 2\\
1 & 2 & 2 & 2\\
1 & 0 & 0 & 0\\
0 & 0 & 1 & 0\\
\end{array}
\right)
\left(
\begin{array}{cccc}
2 & 2 & 2 & 1 \\
2 & 1 & 2 & 2 \\
0 & 2 & 0 & 0 \\
0 & 0 & 0 & 1 \\
\end{array}
\right)
\left(
\begin{array}{cccc}
1 & 2 & 1 & 2 \\
0 & 0 & 0 & 0 \\
2 & 1 & 2 & 1 \\
0 & 0 & 0 & 0 \\
\end{array}
\right)
{\rm for}~ p^{\kappa}=3$ \\
~\\
\hline \end{tabular}
\end{center}
\caption{\small Complete set of independent generators for the non constant harmonic functions for a two-dimensional FES of size $L=4$.}
\label{tab2}
 \end{table}

Let $\rho(m)$ be the number of atoms containing $m$ allowed configurations.
The average occupation  $\bar{m}$ and its standard deviation are then given by:
\begin{equation}
\bar{m}={{\sum_{m} m \rho(m)}\over{\sum_{m} \rho(m)}} ~~~~~~\Delta m={{\sum_{m} (m-\bar{m})^2 \rho(m)}\over {\sum_{m} \rho(m)}}
\end{equation}

\begin{figure}
\begin{center}
\resizebox{0.75\columnwidth}{!}{
\includegraphics{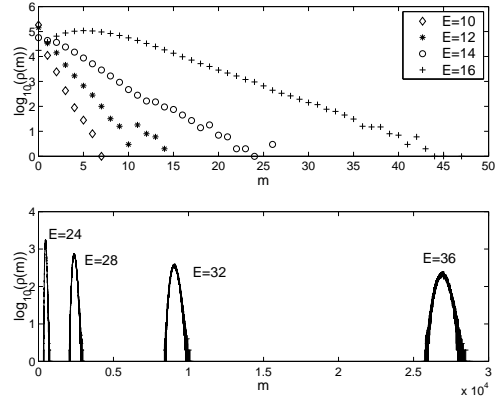}
}
\caption{The logarithm $\rho(m)$ versus $m$ ($\rho(m)$.
The upper panel is relevant to low energies $E$ where $\rho(0)$ is maximum and an exponential decrease
is present. The lower panel refers to higher energies where the occupation is peaked around the average
value $\bar{m}$.}
\label{rho}
\end{center}
\end{figure}

\begin{figure}
\begin{center}
\resizebox{0.75\columnwidth}{!}{
\includegraphics{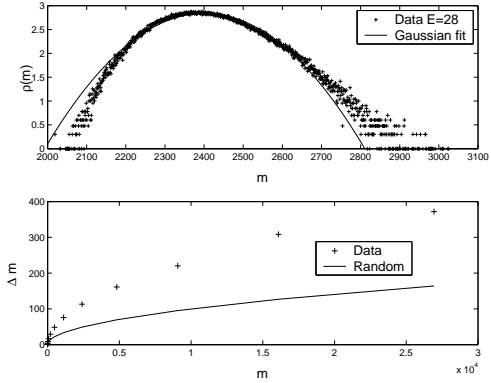}
} \caption{In the upper panel a detailed plot of $\rho(m)$ for 
$E=28$. The comparison with the Gaussian distribution
(same average and  standard deviation) shows an
asymmetric behavior. In the lower panel the standard deviation  
$\Delta m$ vs. the average occupation
number $\bar{m}$. The comparison with the curve $\bar{m}^{1/2}$ shows that
atoms are not filled randomly.} \label{disp}
\end{center}
\end{figure}

The exhaustive calculation for $L=4$ provides $\rho(m)$ at different energies. The results are plotted in Figures \ref{rho} and \ref{disp}.
At very low energies, most of the atoms are empty. In this case they will be denoted as
{\it virtual} atoms, since no permitted configuration exists realizing
the corresponding values of the invariants.
The upper panel of Figure  \ref{rho} shows that for small energies
$\rho(m)$ is maximum at $m=0$, then  it exponentially decreases with $m$.
On the other hand, the lower panel evidences that for higher energies $\rho(m)$ vanishes
at $m=0$, so that  all atoms are filled by allowed configurations. 
In particular, $\rho(m)$ is peaked around the average occupation
$\bar{m}$. Finally, we note that $\bar{m}$ rapidly increases with the energy of the systems.

Let us now analyze the shapes of the peaks. In the upper panel of
Figure \ref{disp}, data calculated for $E=28$ are fitted using a
Gaussian curve with parameters corresponding to the actual average
value $\bar{m}$ and standard deviation $\Delta m$. 
$\rho(m)$ has significant deviations from a Gaussian behavior, with
an asymmetric shape exhibiting a long tail for large occupations. The
lower panel displays the average occupation $\bar{m}$ versus the
corresponding standard deviations $\Delta m$ for different energies.
A random occupation would imply $\Delta m =
\sqrt{\bar{m}}$, while the figure points out a deviation from such
behavior. Precisely, the standard deviations are much larger than
$\sqrt{\bar{m}}$, a signal of correlation in the occupation number.

Since we imposed the constraint $0\leq z_i \leq 7$, 
an analogous transition is present also for $\bar{E}\sim 7$ because of the symmetry
mentioned in Section~\ref{sec2}.
It would be important to establish the robustness of this scenario
at higher $L$, but direct numerical experiments are computationally very demanding.
However, for some of the properties, we have checked that our results are consistent
also at larger sizes.

\section{Low energy transition}\label{sec8}

Assuming that previous results hold independently of the system's size and 
topology, there should exist a transition induced by
the deformation of the space ${\cal C'}$ with respect to ${\cal C}={\bf Z}^{N}$,
affecting only low-(high) energy hyper-surfaces.
To identify the energy of such a crossover, we need to evaluate $C(\bar{E},N)$,
i.e. the number of configurations having energy density $\bar{E}$
on a system of size $N$.
Let $n_i$  be the number of sites having energy $i$ ($i=0,1,\dots, i_{max}$) and $x_i=n_i/N$; then the number of configurations $C(\bar{E},N)$ is
\begin{equation}
C(\bar{E},N)=\int \frac{N!}{\prod_i  N x_i!} d x_i \sim \int e^{-N \sum_i x_i \log(x_i)} d x_i
\end{equation}
In a saddle-point approach, for large $N$, the values of $x_i$ are evaluated by minimizing
$\sum_i x_i \log(x_i)$, with the constraints $\sum_i x_i=1$ and $\sum_i i x_i = \bar{E}$.
By means of Lagrangian multipliers, the minimum of
$\sum_i x_i \log(x_i)$ can be easily obtained in terms of $\bar{E}$ and $i_{max}$.
\begin{equation}
C(\bar{E},L) \sim e^{b_{i_{max}}(\bar{E})N}~.
\label{Nconf}
\end{equation}
Where $b_{i_{max}}(\bar{E})$ is a function that can be analytically evaluated.
By comparing (\ref{count_inv}) and (\ref{Nconf}), whenever $b_{i_{max}}(\bar{E})<c_{th}$
the number of configurations results to be exponentially smaller than the number of atoms,
and most of them do not contain any configuration.
On the other hand, for $b_{i_{max}}(\bar{E})>c_{th}$, all possible atoms are expected to be occupied. 
Therefore, the crossover energy $\bar{E}_1$ can be obtained as $b_{i_{max}}(\bar{E})=c_{th}$,
which gives $\bar{E}_1 \simeq 0.72$.

\begin{figure}
\begin{center}
\resizebox{0.75\columnwidth}{!}{
\includegraphics{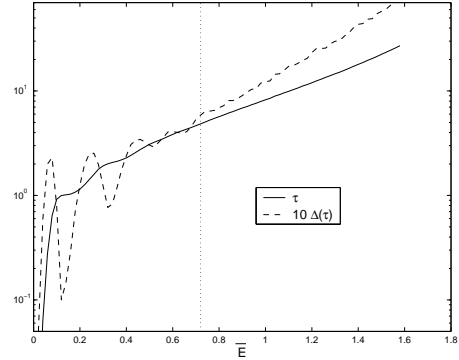}
}
\caption{The average halting time (solid line) and its fluctuations 
(dashed line) vs. the energy density (fluctuation are multiplied by 10,
in order to put plots in the same figure).
The vertical dotted lines denote the crossover density energy $\bar{E}_1$.}
\label{l800}
\end{center}
\end{figure}

Now we want to check if such a crossover has also dynamical consequences.
Since at $\bar{E}_1$ the system is frozen (indeed periodic orbits are present
only for $\bar{E} > \bar{E}_0\simeq 2.06$ \cite{bagnoli}), the only relevant dynamical quantities are the average halting (freezing) time
$\tau$ and its fluctuations $\Delta \tau$.
Now, let $\tau_Z$ be the halting time for the orbit
starting from a configuration $Z$. We define
\begin{equation}
\tau={{\sum_{Z(E)} \tau_{Z(E)}}\over {N_{E}}}~~~~~~~{\Delta \tau}={{\sum_{Z(E)} \tau_{Z(E)}^2}\over
{N_{E}}}-\tau^2
\label{tr}
\end{equation}
where the sums run over the configurations $Z$ at energy $E$, $N_E$ being the
number of such configurations. In the numerical simulations we have averaged
over $5000$ different initial conditions.

\begin{figure}
\begin{center}
\resizebox{0.75\columnwidth}{!}{
\includegraphics{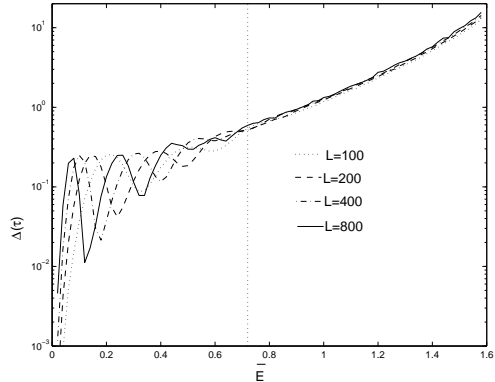}
}
\caption{The fluctuations of the halting time for different lattice sizes.
The vertical dotted lines signal $\bar{E}_1$. Below $\bar{E}_1$, 
fluctuations are oscillating functions of the energy density. For  $\bar{E}>\bar{E}_1$ 
fluctuations increase with the energy and seem to be independent of the
lattice size, corroborating the conjecture about the existence of a transition at $\bar{E}_{1}$.}
\label{DeltaTr}
\end{center}
\end{figure}

In Figure \ref{l800}, $\tau$ and $\Delta \tau$ are plotted versus the
energy density. Since, for such low energies, the halting time
is small, we can perform simulations for very large systems (here, $L=800$).
The vertical dotted line marks the transition between the regions where the number
of atoms is smaller or larger than the number of configurations.
In the former region, $\tau$ is characterized by a series of slow-downs forming small plateaus
where $\tau$ results to be an integer value.
The minima of fluctuations occur in correspondence of these plateaus, while
their maxima are in the transitions between different plateaus.
Such behaviors of $\tau$  and $\Delta \tau$ can be directly explained if
for an energy corresponding to a plateau almost all of the configurations are characterized
by the same halting time. For energies between two different plateaus,
only the two halting times characterizing the near plateaus are possible.
If the dynamics freezes with
equal probability in $n$ or in $n+1$ steps (all other halting times being not allowed),
then $\Delta\tau =1/4$, indeed the height of the fluctuation peaks results to be
about $0.25$.
On the other hand, at energies larger than $\bar{E}_1$, the behaviors of $\tau$
and $\Delta \tau$ are much more regular. This should follow from
the large occupations of the atoms, permitting
many halting times at the same energy. In such case, the system cannot be described
by the simple picture used for $\bar{E}<\bar{E}_1$.

In figure \ref{DeltaTr}, we plot the fluctuations of the halting time as a function of the
energy density for different torus sizes. In the low energy region, by increasing the size of the system oscillations become more and  more rapid and the peaks move towards $\bar{E}=0$.
This is consistent with our picture, since at a given low energy density,
the probability that the evolution freezes into $n$ steps increases with the size. Indeed, this probability depends on the number of way a configuration can contain one or more blocks of 
sites that freeze in a certain time. On the contrary, above the critical energy $\bar{E}_1$, 
the fluctuations seem to be independent of $L$, which is a rather surprising result.

\section{Conclusions}\label{sec9}

In the present work we have investigated the structure of the configuration space
of DFES models on generic unoriented graphs.
Our major result, the complete and algorithmically explicit calculation of the toppling 
invariants, extends to the conservative case
the group theoretical framework introduced by Dhar for dissipative sandpiles.
The Toppling Group seems to be a very general feature linking sandpiles dynamics with 
graph's algebraic properties.
Using this algebraic approach it is possible to identify the exact way the configuration
space is partitioned by the dynamics into invariant subsets, and to determine their
main properties.

The validity of the analytical results is corroborated by an independent numerical analysis
carried out in the case of two-dimensional lattices.

As by-products of our analysis, many properties of the two-dimensional
system are elucidated. 
In particular we give a qualitative explanation for the abundance of orbits with very short
periods. In addition, the further refinement of atoms into attraction's basins reveals 
an unexpected relation with the devil's staircase structure of Ref. ~\cite{bagnoli}. The centres of the plateaus correspond to energies displaying peaks in the number of basins per atom.  
We argue that the absence of a one-to-one correspondence between
periodic orbits and invariant subsets makes the deterministic conservative model
``structurally'' different from both dissipative ones and the stochastic conservative
models, supporting the thesis of a distinct universality class for the deterministic FES. 

We also point out that any further constraint can have relevant
consequences on the organization of the configuration space into invariant sets.
For example, the usual constraint on the energy positiveness introduces a transition
in the frozen phase, saparating the region where most of the atoms are empty from the
region where atoms are filled with a growing number of configurations. 
For small size systems, this analysis can be performed by an exhaustive counting
procedure. On the other hand, for large systems, the transition energy may be analitically
estimated by asymptotic methods. This transition has also a dynamical counterpart 
in the behavior of the halting time.

This work may as well be the basis for a more general study on the
the statistical properties (spectral features, noise etc.)
of stochastic counterparts as dynamical systems presenting a mixture
of randomness and regularity due to the existence of a
partition of the configuration space with an external
weak random perturbation (see \cite{mc2002} for other examples).

Acknowledgment:
One of us (PV) has been supported by a Marie Curie Early Stage Training Programme
fellowship.

\appendix

\section{Modular algebra}\label{appen1}

Since the definition of $K\mathcal{G}$-functionals involves the modulo operation, we
introduce some useful definitions and notations of modular algebra
(for details see e.g. \cite{beachy}).

\begin{definition}
The functions of ${\cal{F}}(\mathcal{G},K)$ are naturally endowed with a structure of
$K$-(Abelian)module (i.e. a vector space where scalars belong to the ring ${\bf Z}_K$
and not to a field).
In particular, the sum of $f$ and $g$ is denoted as $h\=K f+g$ and the product of $f$
and $a\in{\bf Z}_K$ as $h\=K a f$.
$\underline{0}$ is the neutral element of ${\cal{F}}(\mathcal{G},K)$.
\end{definition}

Let ${\cal{M}}(K)$ be a generic $K$-module.
An element $m\in {\cal{M}}(K)$ has periodicity $P\in{\bf Z}_K$, $P<K$, if $Pm\=K
\underline{0}$, in this case $P$ is a factor of $K$.
We shall now proceed in extending as far as possible the basic notion of linear algebra
in the modular sense.

\begin{definition}
A function $F:{\cal{M}}(K)\to {\cal{M}}(K)$ is a
$K$-module homomorphism if for any $m,n\in {\cal{M}}(K)$ and
$a,b\in {{\bf Z}}_K$
$F(a m+b n)\=K a F(m)+b F(n)$. The kernel of $F$ is the set
${\rm Ker}(F)=\{m\in{\cal{M}}(K)| F(m)\=K \underline{0} \}$.
${\rm Ker}(F)$ is a submodule of ${\cal{M}}(K)$. The discrete Laplacian
$ {\widetilde{\nabla}_K^2}$ in Eq.~\ref{dlap} is an example of $K$-module homomorphism
$ {\widetilde{\nabla}_K^2}:{\cal{F}}(L,K)\to{\cal{F}}(L,K)$.
\end{definition}

\begin{definition}
Consider a set $S=\{s_{\alpha}\}\subseteq{\cal{M}}(K)$,
with $\alpha=1,\dots,N$. The elements $\{s_{\alpha}\}$  are algebraically
dependent
if there exists $s_{\beta}\in S$ and a set of $a_{\alpha}\in{\bf Z}_K$
such that
\begin{equation}
s_{\beta} \=K \sum_{\alpha\not=\beta} a_{\alpha} s_{\alpha}
\label{dep}
\end{equation}
Otherwise the elements of $S$ are algebraically independent.
Let $\mathcal{G}\subset {\cal{M}}(K)$ be a set
of algebraically independent elements $\{g_{\alpha}\}$. The element of $\mathcal{G}$
are independent generators, if for any $m \in {\cal{M}}(K)$
there exists a set $a_{\alpha}\in {{\bf Z}}_K$ such that:
\begin{equation}
m \=K \sum_{\alpha} a_{\alpha} g_{\alpha}
\label{dep2}
\end{equation}
A subset $E\subset {\cal{M}}(K)$  of independent generators $\{e_{\alpha}\}$ is a basis if
\begin{equation}
\sum_{\alpha} a_{\alpha} e_{\alpha}=\underline{0}
\label{basis}
\end{equation}
implies $a_{\alpha}\=K 0$ for all $\alpha$. If ${\cal{M}}(K)$ has a basis than
${\cal{M}}(K)$ is called a free $K$-module. We point out that
any $K$-module has a set of independent generators,
while there exist modules where it is not possible to introduce a basis.
\end{definition}

Free $K$-(Abelian)modules have very peculiar properties,
and they are similar for many aspects to the usual vector spaces.
In particular any element  $m\in{\cal{M}}(K)$ can be written in a unique way as
($a_{\alpha}\in {{\bf Z}}_K$):
\begin{equation}
m\=K\sum_{n} a_{\alpha} e_{\alpha}
\label{basis2}
\end{equation}
As a consequence, in a  free $K$-module
any homomorphisms $F$ can be represented by means of matrices.
Let us expand the generic element $m\in {\cal{M}}(K)$ in the basis $\{e_{\alpha}\}$
as in (\ref{basis2}). The components $b_1,b_2,\dots$ of $F(m)$ in the basis $\{e_{\alpha}\}$
result to be:
\begin{equation}
b_{\beta}\=K\sum_{\alpha} F_{\beta,\alpha} a_{\alpha}
\label{homo3}
\end{equation}
where the matrix $F_{\beta,\alpha}$ as in the case of standard vector spaces
is defined by
\begin{equation}
F(e_{\alpha})\=K\sum_{\beta} F_{\beta,\alpha}e_{\beta}
\label{homo2}
\end{equation}

${\cal{F}}({\cal{G}},K)$ is a free module
and the homomorphism $F:{\cal{F}}(L,K)\to{\cal{F}}(L,K)$ is naturally
represented as a matrix in the basis $e_{m}(i)=\delta_{i,m}$.
However, the existence of a basis  for a generic
$K$-module is not guaranteed (the $K$-module is not free).
In particular, some submodules of  ${\cal{F}}({\cal{G}},K)$  are not free.
On a torus, let us call $\chi$ the submodule generated by:
\begin{equation}
g_1=\left(
\begin{array}{cc}
  1  &  0 \\
  0  &  0 \\
\end{array}
 \right)\quad
g_2=\left(
\begin{array}{cc}
  0  &  0 \\
  2  &  0 \\
\end{array}
 \right)
\end{equation}
i.e. any element of $c\in\chi$ is a combination of the form
$c \mathop{=}\limits^4 a_1g_1+a_2g_2$.
$\chi$ is ${{\bf Z}}_4$-module with $g_1$ and
$g_2$ as independent generators. However $\chi$
does not admit any basis.
Hence, a representation of the homomorphisms
$F\chi \to \chi$ in matrix form is not possible.
We note that the element $g_2$ has period $P=2<K=4$.

\section{Theorems 1 and 2}\label{appen2}
In this section, we provide a simple proof of Theorems 1 and 2 for a 
generic undirected
graph ${\cal G}$ with $N$ vertices.
Let us define the sets
\begin{equation}
\label{region} R_k \equiv \{ i : z(i)=k \}~,
\end{equation}
we call {\sl critical} and {\sl metacritical} regions $R_c$ and $R_m$
respectively
\begin{equation}
\label{RCM} R_c \equiv \{ i : z(i)=z_{c}(i) \}~, \quad {\rm and}\quad R_m \equiv \{ i : z(i)>z_{c}(i) \}~.
\end{equation}
The {\it toppling matrix} $T_t$ is defined
by
\begin{equation} \label{matricet} T_t=Z_{t+1}-Z_{t}~. \end{equation}
\begin{theorem}
The  $K{\cal G}$-functional
$\Phi_f(K,{\cal G},Z_t)$  {\em is a toppling invariant if and only
if its generating function $f$ is $K$-harmonic on ${\cal G}$}.
\end{theorem}
\begin{proof}
First we show that $K$-harmonicity is a sufficient condition.
Let ${\bar{ R}_m}(t)$ be the metacritical set $R_m$
at time $t$ indented  with its non metacritical neighbors. Let
${\bar{ R}_m}^c(t)$ be the complementary set, i.e. ${\bar{
R}_m}(t)\cup {\bar{ R}_m}^c(t) = {\cal G}$. At time $t$, the sum
defining $\Phi_f(Z_t)$ in (\ref{functional}) may be split in
two sums running over ${\bar{ R}_m}(t)$ and  ${\bar{ R}_m}^c(t)$
respectively:
\begin{eqnarray}
A_t & \=K \sum_{i=1}^{N} f(i)~z_t(i) & i
\in
{\bar{ R}_m}(t) \\
B_t & \=K \sum_{i=1}^{N} f(i)~z_t(i) & i \in {\bar{ R}_m}^c(t)
\end{eqnarray}
 Similarly we define:
\begin{eqnarray}
{A}_{t+1} &\=K \sum_{i=1}^{N} f(i)~z_{t+1}(i) & i \in
{\bar{ R}_m}(t) \\
{B}_{t+1} &\=K \sum_{i=1}^{N} f(i)~z_{t+1}(i) & i \in {\bar{
R}_m}^c(t)
\end{eqnarray}
Since  ${\bar{ R}_m}^c(t)~,$ is not touched by evolution,
$B_{t+1}=B_t$ and
\begin{equation}\Phi_f(Z_{t+1}) \=K A_{t+1}+B_{t+1} \=K
A_{t+1}+B_t \end{equation}
$A_{t+1}$ may be rewritten as
\begin{equation} A_{t+1}\=K \sum_{i=1}^{N} f(i)~[z_t(i)+T_t(i)],\end{equation}
for $i \in {\bar{ R}_m}(t)$.
Consider now the difference
\begin{equation} \Delta_t \=K A_{t+1}-A_t \=K
\sum_{i=1}^{N} f(i)~T_t(i),\quad i \in {\bar{ R}_m}(t).\end{equation}
The Abelian property of the BTW-rule ensures that the sum
defining $\Delta_t$, can be rewritten as a sum over the strictly metacritical set,
by writing:
\begin{eqnarray}
\Delta_t & \=K & \sum_{i=1}^{N}
\left(-d_{i} f(i)+ \sum_{j \sim i} f(j) \right) \nonumber\\
&\=K & - \sum_{i=1}^{N}  {{\nabla}^2} f(i) \=K 0 \qquad i \in R_m
\end{eqnarray}
where we use the hypothesis of $k$-harmonicity of $f$. Therefore
 \begin{equation}  \Phi_f(Z_{t+1}) \=K
 A_t+B_t+\Delta_t
 \=K A_t+B_t   \=K \Phi_f(Z_t) \nonumber \end{equation}
 leading to the
conclusion that $K$-harmonicity implies toppling invariance.

Now we prove that $K$-harmonicity is a necessary condition. The
hypothesis is that, for every $Z_t$
\begin{equation} \Phi_f(Z_{t+1}) \=K \Phi_f(Z_t)\end{equation}
and the claim is $( {{\nabla}^2} f)(m)\=K 0$, for an arbitrary $m$.
Let in particular $\bar{Z_t}(i)=d_{i}\delta_{i,m}$,
then, $\bar{Z}_{t+1}(i)= \sum_{j \sim m}\delta_{i, j}$.
This means
\begin{equation}
 \Phi_f(\bar{Z}_{t+1})- \Phi_f(\bar{Z_t}) \=K ( {{\nabla}^2} f)(m) \=K 0 ~,
\end{equation}
\qed.
\end{proof}

\begin{theorem}\label{th_iso}
A necessary and sufficient condition for a transformation to be
isoinvariant is that it can be written in the form
\begin{equation}
Z' = \eta(Z) = Z + {\nabla}^2 U
\end{equation}
where $U \in {\bf Z}^{N}$ is an integer vector.
\end{theorem}
\begin{proof}
Two configuration related by formula (\ref{iso}) are clearly isoinvariant
since
\begin{eqnarray}
\Phi_h(K,{\cal G},Z')& \=K & \sum_{i=1}^{N} h(i) \left(Z(i)
+({{\nabla}^2} U)(i)\right) \nonumber\\
~ & \=K & \sum_{i=1}^{N} h(i) Z(i)\nonumber + \sum_{i=1}^{N}
({{\nabla}^2} h)(i)U(i) \nonumber\\
~ & \=K &  \Phi_h(K,{\cal G},Z)
\label{iso2}
\end{eqnarray}
where we used the symmetry properties of ${\nabla}^2$
and the fact that $h$ is a $K$-harmonic function.

Let $W=Z'-Z$ be the difference between the configurations $Z$ and $Z'$ 
belonging to the same atom,
we have that $B^{\dag} W = SL$, where $B$ and $S$ are the unimodolar and the 
diagonal matrices appearing in the Smith decomposition (\ref{smith}), 
and $L$ is an integer vector.
The Smith decomposition yields
$B^{\dag} W =
B^{\dag}\nabla^2 A^{\dag}L$
and then $W= {\nabla}^2 A^{\dag} L$, 
therefore $W$ can be obtained applying the Laplacian matrix to the integer
vector $A^{\dag} L$ \qed.
\end{proof}

\section{Independence in ${\cal H}({\cal G},K)$}\label{appen3}

The algebraic independence (\ref{dep}) between functions of the $K$-module
${\cal H}({\cal G},K)$ implies that they are also independent according to
definition (\ref{dep_inv}). This is a simple consequence of the fact
that the toppling invariants are ``linear functional''
in the space ${\cal F}({\cal G},K)$.
Therefore the partition of the energy surface
generated by the whole set of $K$-harmonic functions ${\cal{H}}({\cal G},K)$
is the same as the partition generated by a subset
of independent generators of ${\cal{H}}({\cal G},K)$.
On the contrary of  ${\cal{F}}({\cal G},K)$,  ${\cal{H}}({\cal G},K)$ is not in general a
free module (\ref{basis}). In particular there may be  generators
of periodicity $P<K$.

Let us now consider the harmonic functions corresponding to different
$K$'s. Since only the constant functions
belong to ${\cal{H}}({\cal G},K)$ for any  $K$, the energy
is a special invariant to be considered apart, as in definition (\ref{dep_inv}).
To this purpose we denote with $\mathcal{H'}({\cal G},K)$ any set of generators
of ${\cal{H}}({\cal G},K)$ such that  $c(i)=1, \forall i$ belongs
to $\mathcal{H'}({\cal G},K)$.
The following theorem shows that the partition of the energy surface generated
by all the $K$-harmonic functions can be evaluated by considering only
$K$'s of the form $K=p^{\kappa}$, where $p$  is a prime number and $\kappa$ is an
integer depending on $p$, i.e. considering the invariant generated by the functions
of ${\cal H'}({\cal G},p^{\kappa})$.
Let us first introduce a lemma which is a basic properties of integer number:

\begin{lemma}
Let $K=p_1^{{\kappa}_1}p_2^{{\kappa}_2}\dots p_r^{{\kappa}_r}$
the decomposition of $K$ into prime factors $p_n$. Any element of $b\in {{\bf Z}}_K$
 can be written in a unique way as:
\begin{equation}
b\=K\sum_{n=1}^r q_n b_n
\label{decomp}
\end{equation}
where $b_n\in{{\bf Z}}_{\{p_n^{{\kappa}_n}\}}=\{0,1,\dots,p_n^{{\kappa}_n}-1\}$;
and the integers $q_n$ are given by:
\begin{equation}
q_n=\prod_{p_h\not=p_n}p_h^{{\kappa}_h}
\label{defq}
\end{equation}
\end{lemma}

\begin{theorem}
Let us consider the module ${\cal{H}}({\cal G},K)$,
and let $K=p_1^{{\kappa}_1}p_2^{{\kappa}_2}\dots p_r^{{\kappa}_r}$ the decomposition of
$K$ into prime factors $p_n$.
For any $h\in{\cal{H}}({\cal G},K)$ there is a set of functions
$h_n\in{\cal{H}}({\cal G},p_n^{{\kappa}_n})$ such that $h$ depends on the $h_n$'s according
to definition (\ref{dep_inv}).
\end{theorem}
\begin{proof}
\begin{sloppypar}
>From Lemma 1, any elements $h(i)$ of the function (matrix in two dimensions) $h\in{\cal{H}}({\cal G},K)$ can be
decomposed in a unique way as:
\begin{equation}
h(i)\=K\sum_{n=1}^r q_n h_n(i)
\label{factor1}
\end{equation}
where $h_n(i)\in{{\bf Z}}_{\{p^{{\kappa}_n}\}}=\{0,1,\dots,p_n^{{\kappa}_j}-1\}$.
Equation (\ref{factor1}) defines the functions $h_n$'s.
Indeed we have
\begin{equation}
\underline{0}_K \=K  {\widetilde{\nabla}^2_K}(h)\=K\sum_{n=1}^r q_n  {\widetilde{\nabla}^2_K}(h_n)
\label{factor2}
\end{equation}
Lemma 1 entails that the neutral element $\underline{0}_K$ of ${\cal{H}}({\cal G},K)$
can be obtained, by means of factorization (\ref{factor1}),
only as a combination of the neutral elements
$\underline{0}_{\{p_n^{{\kappa}_n}\}}$ of ${\cal{H}}({\cal G},p_n^{{\kappa}_n})$.
Therefore, Equation (\ref{factor2}) means that, for all $n$,
$h_n\in{\cal{H}}({\cal G},p_n^{{\kappa}_n})$.
Finally for the $K{\cal G}$-functionals we have
\begin{eqnarray}
\Phi_h(K,{\cal G},Z)& \=K & \sum_{n=1}^r q_n\left(\sum_{i=1}^{N} h_n(i)~z(i)\right)\nonumber\\
&\=K&
 \sum_{n=1}^r q_n \Phi_{h_n}(p_n^{{\kappa}_n},{\cal G},Z)
\label{factor4}
\end{eqnarray}
in the last expression we used the equality $q\cdot a ~\=K ~q \cdot(a~{\rm mod}~(K/q))$
which holds when $K/q$ is an integer, (here,  from (\ref{defq}), $K/q=p_n^{{\kappa}_n}$).
\qed.
\end{sloppypar}
\end{proof}

We now introduce a theorem which allows to simplify the study of 
${\cal{H}}({\cal G},p^\iota)$ for different values of the integer
$\iota$. The following Lemma can be
directly proved resorting to the properties of modular spaces.

\begin{lemma}
A periodic element  $h\in{\cal{H}}({\cal G},p^{\kappa})$ of periodicity
$p^\iota$, i.e. $p^{\iota} h(i) \pka 0$,
can be written as $h(i)=p^{\kappa-\iota} h'(i)$
with  $h'\in{\cal{H}}({\cal G},p^{\iota})$.
\end{lemma}

\begin{theorem}
Let us consider an integer $\kappa$ such that the any function $h\in {\cal{H}}({\cal G},p^{\kappa})$
can be written as  $ h(i)=h'(i)+c$ where $c$ is a constant and $h'$ is a function
of periodicity smaller than $p^{\kappa}$.
Any function  belonging to ${\cal H}({\cal G},p^{\epsilon})$, for any integer $\epsilon$,
is dependent on a function $h\in {\cal{H}}({\cal G},p^{\kappa})$.
\end{theorem}
\begin{proof}
\begin{sloppypar}
A generic function  belonging to  ${\cal{H}}({\cal G},p^{\epsilon})$
will be denoted as $h^{(\epsilon)}$.
First we focus on the case $\epsilon<\kappa$.
For any element  $h^{(\epsilon)}\in {\cal{H}}({\cal G},p^{\epsilon})$ we consider
the element $h'\in{\cal{F}}({\cal G},p^{\kappa})$, given by $h'(i)=p^{\kappa-\epsilon} h(i)$.
Since
$( {{\nabla}^2} h')(i) \ph p^{h-\epsilon} ( {{\nabla}^2} h)(i)
\ph  p^{h-\epsilon}(a~ p^\epsilon)\ph 0$
($a$ is an integer), we have
$h'\in{\cal{H}}({\cal G},p^{h})$. Moreover,
$\Phi_{h^{(\epsilon)}}(p^{\epsilon},{\cal G},Z)=
p^{\epsilon-\kappa} \Phi_{h'}(p^{\kappa},{\cal G},Z)$
and this concludes the first part of the proof.
\end{sloppypar}
\begin{sloppypar}
We pass now to the case $\epsilon \geq \kappa$.
If all the functions $h^{(\kappa)}\in {\cal{H}}({\cal G},p^{\kappa})$ are a
combination of a constant
and a function of periodicity smaller than $p^{\kappa}$,
Lemma 2 implies that
 \begin{equation}
\label{h(k)}
 h^{(\kappa)}(i)=p h^{(\kappa-1)}(i)+c
\end{equation}
with $h^{(k-1)}\in {\cal{H}}({\cal G},p^{\kappa-1})$.
First we show by induction that in any module
${\cal{H}}({\cal G},p^{\epsilon})$ with $\epsilon \geq \kappa$
all the functions $h^{(\epsilon)}$ are a combination
of a constant function and of a
periodic harmonic function of period smaller than $p^\kappa$, i.e.
from Lemma 2
\begin{equation}
\label{h(epsilon)}
h^{(\epsilon)}(i)=p^{\epsilon-\kappa +1} h^{(\kappa-1)}(i)+b
\end{equation}
In the case $\epsilon=\kappa$
this property is verified by hypothesis. We now suppose that the property holds
for $h^{(\iota-1)}$ and show that this is true also for $h^{\iota}$.
We define
$h^{*}(i) \pio p^{\iota-\kappa} h^{(\iota)}(i).$
Since $p^{\kappa}h^{*}\pio 0$, from Lemma 2 we have that
$h^{*}= p^{\iota-\kappa} h^{(\kappa)}$ with $h^{(\kappa)}\in{\cal{H}}({\cal G},p^{\kappa})$.
We have
$p^{\iota-\kappa} h^{(\iota)}(i)~\pio ~~h^{*}(i)\pio p^{\iota-\kappa} h^{(\kappa)}(i)$
and multiplying both sides by $p^{\kappa-1}$ we get
$p^{\iota-1} h^{(\iota)}(i)~ \pio ~ p^{\iota-1} h^{(\kappa)}(i)$. Now $h^{(\kappa)}$ satisfies
(\ref{h(k)}) and therefore
\begin{equation}
p^{\iota-1} h^{(\iota)}(i) \pio p^{\iota-1} (p h^{(\kappa-1)}(i)+c) \pio  p^{\iota-1} c
\label{h(iota)}
\end{equation}
Equation (\ref{h(iota)}) proves that $h^{(\iota)}(i)-c$ is periodic of period
$p^{\iota-1}$ and then from Lemma 2
$h^{(\iota)}(i)-c=p  h^{(\iota-1)}(i)$
where $h^{(\iota-1)}(i)\in {\cal{H}}({\cal G},p^{\iota-1})$. Since by induction
we supposed that Equation (\ref{h(epsilon)}) holds when $\epsilon=\iota-1$,
we get $h^{(\iota)}(i)-c=p (p^{\iota-\kappa} h^{(k-1)}(i)+b)$
i.e. $h^{(\iota)}(i)= p^{\iota-\kappa +1} h^{(k-1)}(i)+d $
where $d$ is an integer constant. This concludes the prove by induction.
Finally, we consider the toppling invariants
$\Phi_{h^{\epsilon}}(p^{\epsilon},{\cal G},Z)$, from (\ref{h(epsilon)})
we have $\Phi_h(p^{\epsilon},{\cal G},Z) \pie
p^{\epsilon-\kappa+1} \Phi_{h^{(\kappa-1)}}(p^{\kappa-1},{\cal G},Z)+ d E(Z)$.  
\qed
\end{sloppypar}
\end{proof}

We note that Theorem 4 does not prove that the integer $\kappa$ exists. 
This can be proved, for instance by means of Smith decomposition 
(Section \ref{sec4}).

\section{Algorithm }\label{appen4}

\subsection{Matrix equations}\label{subappen4a}

Let us consider a free module ${\cal{M}}(K)$ and its basis
$e_{\alpha}$. Since any
homomorphism $F:{\cal{M}}(K)\to{\cal{M}}(K)$ is represented by a
matrix $F_{\beta,\alpha}$, the equations determining the components $k_{\alpha}$
of the elements of ${\rm Ker}(F)$ are
\begin{equation}
\sum_{\alpha=1}^M F_{\beta,\alpha}~ k_{\alpha} \=K 0
\label{ker}
\end{equation}
where $M$ is the number of elements of the basis.
Let us show that the system (\ref{ker}), can be faced
by resorting to techniques similar to those usually adopted in
ordinary vector spaces. First,
the following operations do not change the number of solutions of (\ref{ker}).

\begin{itemize}

\item {\it Substitution.} If the element $F_{\beta,\gamma}$ of the matrix defining system
(\ref{ker}) equals to $1$,  the corresponding element $k_{\gamma}$
can be eliminated by the substitution
\begin{equation}
k_{\gamma} \=K - \sum_{\alpha\not=\gamma} F_{\beta,\alpha}~ k_{\alpha}
\label{ker2}
\end{equation}
Equation (\ref{ker2}) can be discarded reducing
the system from $M$ to $M-1$ equations and variables.

\item {\it Columns and rows exchange.} It is possible to swap two rows or two
columns of the
matrix $F_{\beta,\alpha}$. This corresponds to a relabeling of
equations and variables.

\item {\it Row sum and difference.} 
The matrix elements $F_{\beta,\alpha}$ in the row $\beta$ can be substituted
with $F_{\beta,\alpha}\pm a F_{\gamma,\alpha}$ where $\gamma$ is a row different
from $\beta$ and $a$ is an element of ${\bf Z}_K$.

\item {\it Row multiplication.} Let $a\in{\bf Z}_K$.
If $a$ is not a factor of $K$, then all the matrix elements $F_{\beta,\alpha}$ 
corresponding to the row $\beta$
can be multiplied by $a$.

\end{itemize}

On a $L\times L$ torus, a basis for  ${\cal{F}}({\cal G},K)$ is
$e_{m,n}(i_1,i_2)=\delta_{i_1,m}\delta_{i_2,n}$. The homomorphism corresponding to
the discrete Laplacian $ {\widetilde{\nabla}^2_K}$ can be represented as a $L^2\times L^2$ matrix.
Before introducing the algorithm, we show that the 
system determining ${\rm Ker}( {\widetilde{\nabla}^2_K})$ can be
reduced by substitution from  $L^2$ to $2L$ variables. The matrix form of the harmonic
function $h\in {\cal{H}}({\cal G},K)$ is
\begin{equation}
\label{matrixh}
h=\left(
\begin{array}{cccc}
  h(0,0)  & h(1,0)   &  \cdots  &  h(L-1,0)\\
  h(0,1)  & h(1,1)   &  \cdots  &  h(L-1,1)\\
  h(0,2)  & h(1,2)   &  \cdots  &  h(L-1,2)\\
  \vdots  &  \vdots  &  \ddots  &  \vdots \\
  h(0,L-1)  & h(1,L-1)   &  \cdots  &  h(L-1,L-1)\\
\end{array}
 \right)
\end{equation}
Equation  (\ref{dKh}) yields $h(j,2)\=K 4h(j,1)-h(j-1,1)-h(j+1,1)-h(j,0)$.
In the same way the elements of the fourth row are determined by the elements of third
and second row, and so on. This way the system $ {\widetilde{\nabla}^2_K}(h)\=K0$ is reduced
to a system of $2L$ unknown variables only, i.e. the elements of the two first rows
of (\ref{matrixh}).
This reduction procedure can be implemented into a computer program
allowing to study the homomorphism kernels even for very large systems.

\subsection{Triangularization procedure}\label{subappen4b}

First of all, we recall that
${\bf Z}_{p^{\kappa}}$ is not a field, therefore we
have to pay attention to some algebraic details.
Any element $a\in{\bf Z}_{p^{\kappa}}$ can be decomposed as $a=p^h q$
with $h<{\kappa}$ and $q$ an integer such that $q~ {\rm mod}~p \not= 0$ (i.e.
$q$ admits the reciprocal $q^{-1}\in{\bf Z}_{p^{\kappa}}$ ), we
will call this product $p$-factorization.
Moreover, given two elements $a$ and $a'$ of ${\bf Z}_{p^{\kappa}}$
with $p$-factorizations $a=p^h q$ and $a'=p^{h'} q'$,
if $b\pka a+a'$ and  $c=p^{h_c}q_c$, we have that
$b=p^{h_b}q_b$ with $h_b\geq {\rm min}(h,h')$, and  $c=p^{h_c}q_c$ with
$h_c=h'+h'$ (if $(h+h')\geq {\kappa}$ then $c\pka 0$).
Taking into account these properties of ${\bf Z}_{p^{\kappa}}$,
the matrix $F$ in equation (\ref{ker})
can be put in a triangular form a.s follows

\begin{itemize}

\item Consider the $p$-factorization of all the matrix elements
$F_{\beta,\alpha}=p^{h_{\beta,\alpha}}q_{\beta,\alpha}$ and find
the elements $F_{\beta,\alpha}$ with the smallest exponent $h_{\beta,\alpha}$.
Swap rows and columns so that one of these elements is placed in the position
$(1,1)$ of the transformed matrix $F^A$. 
Then $F_{\beta,\alpha}^A=p^{h_{\beta,\alpha}^A}q_{\beta,\alpha}^A$ with
$h_{1,1}^A\leq h_{\beta,\alpha}^A$ for all $\beta$ and $\alpha$.

\item Multiply the first row of $F_{\beta,\alpha}^A$ by $(q_{\beta,\alpha}^A)^{-1}$.
We get  $F_{\beta,\alpha}^B=p^{h_{\beta,\alpha}^B}q_{\beta,\alpha}^B$
with  $q_{1,1}^B=1$ and
$h_{1,1}^B\leq h_{\beta,\alpha}^B ~ \forall \beta,\alpha$. 

\item \begin{sloppypar}
Subtract to all the elements $p^{h_{\beta,\alpha}^B}q_{\beta,\alpha}^B$
($\beta\not= 1$)
the elements of the first row
multiplied by $p^{(h_{\beta,1}-h_{1,1})}q_{\beta,1}$;
i.e. transform $F^B$ into the matrix 
$F^C_{\beta,\alpha}\pka p^{h_{\beta,\alpha}^B}q_{\beta,\alpha}^B $ if $\beta=1$
and $F^C_{\beta,\alpha}\pka p^{h_{\beta,\alpha}^B}q_{\beta,\alpha}^B
-p^{h_{\beta,1}^B-h_{1,1}^B+h_{1,\alpha}^B}q_{\beta,1}^Bq_{1,\alpha}^B $
if $\beta \not=1$. We have
\def\tempa{\multicolumn{1}{|c}{F^1}}
\begin{equation}
F^C=\left(
\begin{array}{cc}
  p^{h_{1,1}}  & F_{1,2}^C  \cdots F_{1,M}^C \\ \cline{2-2}
  {{{\displaystyle 0} \atop {\displaystyle \vdots}}\atop{{\displaystyle ~} \atop {\displaystyle 0}} }   & \tempa    \\
\end{array}
 \right)
\end{equation}
where $F^1$ is a matrix of size $(M-1)\times (M-1)$. As a consequence of the properties
of sums and products of $p$-factorizations, 
$F_{\beta,\alpha}^1=p^{h_{\beta,\alpha}^1}q_{\beta,\alpha}^1$
with $h_{\beta,\alpha}^1\geq h_{1,1}$. The same property holds for $F_{1,2}^C,\dots,
F_{1,M}^C$.
\end{sloppypar}

\item If all elements of $F^1$ equals zero the triangularization is concluded. Otherwise,
apply to $F^1$ the same procedure applied to $F$. In this second case,
we obtain a transformed matrix whose upper left element is given by $p^{h_{2,2}}$
with $h_{2,2}\geq h_{1,1}$.

\end{itemize}

The algorithm described above transform the general matrix $F$, into a triangular matrix of the form:
\begin{equation}
F^E=\label{UpTPk}
\left(
\begin{array}{c|c|c|c}
T^0& E^{0,1} & \cdots & E^{0,k} \\
\hline
O^{1,0} & T^1 & \cdots & E^{1,{\kappa}} \\
\hline
\vdots & \vdots & \ddots & \vdots \\
\hline
O^{{\kappa},0} & O^{{\kappa},1} & \cdots & T^{\kappa}
\end{array}
\right)
\end{equation}
where $T^h$ ($0\leq h \leq {\kappa}$) are upper-triangular square
matrices of size $Q_{p^h}\times Q_{p^h}$, i.e.
\begin{equation}
\label{UpTPk2}
T^h= \left(
\begin{array}{cccc}
p^h &  T^h_{1,2} & \cdots &  T^h_{1,Q_{p^h}} \\
0 &  p^h & \cdots &  T^h_{2,Q_{p^h}} \\
\vdots & \vdots & \ddots & \vdots \\
0 & 0 & \cdots & p^h
\end{array}
\right)
\end{equation}
therefore we have $\sum_{h=0}^{\kappa} Q_{p^h}=M$ where $M\times M$ is the size
of the original matrix $F$.
For $T^h$ we have
$T^h_{\beta,\alpha}=p^{m_{\beta,\alpha}^{T^h}}q_{\beta,\alpha}^{T^h}$ with
$m_{\beta,\alpha}^{T^h}\geq h$.
All the elements of the $Q_{p^h}\times Q_{p^j}$ rectangular matrices $O^{h,j}$ equal zero.
While, for the $Q_{p^h}\times Q_{p^j}$ matrices $E^{h,j}$, 
$E^{h,j}_{\beta,\alpha}=p^{m_{\beta,\alpha}^{E^{h,j}}}q_{\beta,\alpha}^{E^{h,j}}$,
with  $m_{\beta,\alpha}^{E^{h,j}}\geq h$.
If ${\kappa}=1$, $K$ is a prime, $\cal{M}(K)$ is a vector space and not a module and we recover the usual Gauss elimination procedure,
in particular, all the elements have the reciprocal and we can discard
the $p$-factorization.
By substitution, one can directly check that, for each $h$, $1\leq h \leq \kappa$,
equation $\sum_{\alpha} F_{,\beta,\alpha} k_{\alpha} \pka 0$ has $Q_{p^h}$ independent
solutions of periodicity $p^h$, which generate the submodule ${\rm Ker}(F)$.

The triangularization procedure for the discrete Laplacian
can be directly implemented in a computer program. Main limitations consist
in the computer capability in operations with integers and modules.
For example, since one has to evaluate quantities such
as $(a\cdot b)~{\rm mod}~p^\kappa$,
$p^{2 \kappa}$ has to be smaller than the maximum integer. Furthermore,
the  ${\rm mod}~p^\kappa$ operation turns out to be very slow
for large $p^\kappa$, limiting our calculations to values of $p^\kappa$ up to
20000. On the other hand, the reduction from $L^2$ variable to
$2L$ variables, as described in the previous section,  allows to study
systems of a quite large size. As an instance, for a small value of $p^\kappa$
(i.e. $p=103$, $\kappa=1$),
the kernel of $ {\widetilde{\nabla}^2_{p^\kappa}}$ on a torus of size $L=400$ can be
calculated in a few seconds on a standard personal computer.

The algorithm has been implemented in the following way. We considered all the primes $p$
smaller than $K_L=20000$ and we apply the algorithm to the operator
$ {\widetilde{\nabla}^2_K}$ for $K=p,p^2\dots$ up to the value $p^\kappa$ such that in the diagonalization
procedure we get only one function of periodicity $p^\kappa$ i.e. the constant.
In general we find that $p^\kappa$ is smaller than $K_L$ at least for not too large sizes.
Theorem 3 proves that it is useless considering larger $p^\iota$.
A simple analysis of the diagonal elements
of the triangular matrix (\ref{UpTPk}) allows to evaluate how many elements of
periodicity $p^\iota<p^\kappa$ belongs to $I_{{\cal G},K_{\cal G}}$. 
By Lemma 2 such functions
can then be written as elements of
${\cal H}({\cal G},p^\iota)$.

\section{Preflows and flows on graph}\label{appen5}

\begin{definition}
Given an oriented graph $\mathcal{G}$, we call integer ``preflow''  $P_{(i,j)}$ an integer-valued function $\{(i,j)\} \to {\bf Z}^{N_E}$, where $N_E$ is the number of edges. 
The preflows constitutes a group $\cal {P}$ with the addition in ${\bf Z}$.
\end{definition}

We expose here some properties of the integer preflow group ${\cal P}$
providing a interesting bridge between isoinvariant
transformations and the graph topology. 
\begin{definition}\label{isoen}
We denotes with $\{\tau\}_{en}$ the class of isoenergetic transformations
$\tau: Z\in{\cal C}\to Z'\in{\cal C}$ defined as $Z'=Z+\tau$ where
$\tau\in {\bf Z}^N$ is an integer vector such that $\sum_i \tau_i=0$.
Clearly $\{\tau\}_{en}$ forms an Abelian group.
\end{definition}

The isoinvariant transformations (\ref{iso}) $\{\eta\}_{iso}$ are
a subgroup of $\{\tau\}_{en}$. Moreover, we note that once 
it is fixed the state $Z$, Definition \ref{isoen} provides a one to one
correspondence between the states of an energy surface and the
elements of the group $\{\tau\}_{en}$. In the same way the configurations
of an atom are in a one to one correspondence with the isoinvariant
transformations $\{\eta\}_{iso}$. Therefore, the number of atoms
${\cal N}({\cal G})$ can be evaluated as $|\{\tau\}_{en}/ \{\eta\}_{iso}|$
where $|\cdot|$ denotes the order of the quotient group.

\begin{definition}\label{morph}
We denote with $\Xi$ the morphism associating
to each integer preflow $P_{(i,j)}\in{\cal P}$ the isoenergetic 
transformation $\tau=\Xi(P_{(i,j)}): Z\in{\cal C}\to Z' \in {\cal C}$,  
$Z_i'=Z_i+\sum_j P_{(i,j)}$. We note that ${\rm Range}(\Xi)=\{\tau\}_{en}$
while we denote as ${\cal F}={\rm Ker}(\Xi)$ the subgroup
of integer(conservative) flows, i.e. $P_{(i,j)}\in {\cal F}$  
if $\sum_j P_{(i,j)}=0$.
\end{definition}
\begin{theorem}\label{irrot}
$\Xi(P_{(i,j)})$ is isoinvariant if and only if $P_{(i,j)}$ belongs 
to the subgroup $\cal I$ of the irrotational elements of ${\cal P}$, i.e. if for
each loop of the graph $i_0,i_1,\dots,i_n,i_0$ ($i_k$ and $i_{k+1}$
are adjacent sites) $P_{(i_0,i_1)}+P_{(i_1,i_2)}+\dots+P_{(i_n,i_0)}=0$.
\end{theorem}
\begin{proof}
If $\Xi(P_{(i,j)})$ is isoinvariant we have $\sum_j P_{(i,j)}=(\nabla^2U)_i$
for a certain integer vector $U$. This means that
$P_{(i,j)}=U_i-U_j$ yielding $P_{(i,j)}\in {\cal I}$.
Let $P_{(i,j)}\in {\cal I}$,
the integer vector $U$ can be defined as $U_1=0$ and 
$U_i=P_{(1,i_1)}+P_{(i_1,i_2)}+\dots+P_{(i_{n},i)}$
($0,i_1,\dots,i_n,i$ is a path joining the vertices $0$ and $i$),
the result does not depend on the choice of the path since $P_{(i,j)}$
is irrotational. We get $\sum_j P_{(i,j)}=(\nabla^2U)_i$ therefore
$\Xi(P_{(i,j}))$ is isoinvariant.\qed
\end{proof}

We define ${\cal P}^\#$ as the dual space of ${\cal P}$ i.e.
the set of all linear functions ${\cal P}\to{\rm Z}$. 
${\cal P}^\#$ is isomorphic to ${\cal P}$, indeed any linear functional
can be written in a unique way as $\sum_{(i,j)} P^\#_{(i,j)} P_{(i,j)}$
with $P^\#_{(i,j)}\in{\cal P}$.
\begin{lemma}\label{dual}
The dual space ${\cal F}^\#$ is isomorphic to ${\cal P}/{\cal I}$.
\end{lemma}
\begin{proof}
\begin{sloppypar}
The result follows showing that when $P_{(i,j)}\in{\cal F}$ then
$\sum_{(i,j)} P^\#_{(i,j)} P_{(i,j)}=0$ if and only if
$P^\#_{(i,j)}\in{\cal I}$.  \qed
\end{sloppypar}
\end{proof}
\begin{theorem}\label{complex}
The number of atoms generated by toppling invariants
equals the graph complexity.
\end{theorem}
\begin{proof}
>From Definition \ref{morph} we get that $\{\tau\}_{en}$ is isomorphic
to ${\cal P}/{\cal F}$. Since Theorem \ref{irrot} shows that 
$\cal I$ is isomorphic to $\{\eta\}_{iso}$ we get that
${\cal N}({\cal G})=|\{\tau\}_{en}/ \{\eta\}_{iso}|=
|({\cal P}/{\cal F})/{\cal I}|=|({\cal P}/{\cal I})/{\cal F}|$.
Finally an important result proved in \cite{bacher} shows that
the graph complexity equals $|{\cal F}^\#/{\cal F}|$ and then
from lemma \ref{dual} $|({\cal P}/{\cal I})/{\cal F}|$ and this
completes the proof. \qed
\end{proof}

\end{document}